\documentstyle[12pt]{article}
\begin{document}
\large
\bibliographystyle{plain}
\begin{titlepage}
\large
\hfill\begin{tabular}{l}HEPHY-PUB 652/96\\ UWThPh-1996-48\\ August 1996
\end{tabular}\\[2.5cm]
\begin{center}
{\Large\bf SPINLESS SALPETER EQUATION:}\\[.5ex]
{\Large\bf LAGUERRE BOUNDS ON ENERGY LEVELS}\\
\vspace{2cm}
{\Large\bf Wolfgang LUCHA}\\[.5cm]
Institut f\"ur Hochenergiephysik,\\
\"Osterreichische Akademie der Wissenschaften,\\
Nikolsdorfergasse 18, A-1050 Wien, Austria\\[2cm]
{\Large\bf Franz F.~SCH\"OBERL}\\[.5cm]
Institut f\"ur Theoretische Physik,\\
Universit\"at Wien,\\
Boltzmanngasse 5, A-1090 Wien, Austria\\[2cm]
{\bf Abstract}
\end{center}
\normalsize
\noindent
The spinless Salpeter equation may be considered either as a standard
approximation to the Bethe--Salpeter formalism, designed for the description
of bound states within a relativistic quantum field theory, or as the most
simple, to a certain extent relativistic generalization of the costumary
nonrelativistic Schr\"odinger formalism. Because of the presence of the
rather difficult-to-handle square-root operator of the relativistic kinetic
energy in the corresponding Hamiltonian, very frequently the corresponding
(discrete) spectrum of energy eigenvalues cannot be determined analytically.
Therefore, we show how to calculate, by some clever choice of basis vectors
in the Hilbert space of solutions, for the rather large class of power-law
potentials, at least (sometimes excellent!) upper bounds on these energy
eigenvalues, for the lowest-lying levels this even analytically.

\vspace*{1ex}

\large
\end{titlepage}

\section{Introduction: The Spinless Salpeter Equation}

One's attitude to the well-known ``spinless Salpeter equation'' may be
reflected by either of the following two approaches (or points of view):
\begin{itemize}
\item On the one hand, this spinless Salpeter equation may be regarded to
represent some standard approximation to the Bethe--Salpeter formalism for
the description of bound states within a relativistic quantum field theory.
It may be derived from the Bethe--Salpeter equation \cite{salpeter51} by two
steps:
\begin{enumerate}
\item Eliminate---in full accordance with the spirit of instantaneous
interactions---any dependence on timelike variables to obtain in this way the
so-called Salpeter equation \cite{salpeter52}.
\item Neglect any reference to all the spin degrees of freedom of the
involved bound-state constituents and restrict your formalism exclusively to
positive-energy solutions.
\end{enumerate}
\item On the other hand, this spinless Salpeter equation may be viewed as one
of the most straightforward generalizations of the standard nonrelativistic
quantum theory towards the reconciliation with all the requirements imposed
by special relativity. To be precise, this generalization consists of
incorporating the square-root operator of the relativistic expression for the
kinetic energy of the involved particles. For the particular case of two
particles of equal mass $m$ and relative momentum ${\bf p}$, the
kinetic-energy operator $T$ is given by
\begin{equation}
T({\bf p}) \equiv 2\,\sqrt{{\bf p}^2 + m^2}\ .
\label{eq:kinenergy}
\end{equation}
All the forces operating between the bound-state constituents are tacitly
assumed to be described by an arbitrary static interaction potential $V$. For
the special case of two particles, this interaction potential should depend
only on the relative coordinate ${\bf x}$ of these particles: $V = V({\bf
x})$.
\end{itemize}
In any case, the self-adjoint Hamiltonian $H$ governing the dynamics of any
quantum system to be described by the spinless Salpeter equation will be of
the form
\begin{equation}
H = T({\bf p}) + V({\bf x})\ .
\label{eq:ham-sseq}
\end{equation}
The two-particle spinless Salpeter equation to be investigated here is then
nothing else but the eigenvalue problem for this Hamiltonian $H$,
$$
H|\chi_k\rangle = E_k|\chi_k\rangle\ ,\quad k = 0,1,2,\dots\ ,
$$
for Hilbert-space eigenvectors $|\chi_k\rangle$ corresponding to energy
eigenvalues
$$
E_k \equiv
\frac{\langle\chi_k|H|\chi_k\rangle}{\langle\chi_k|\chi_k\rangle}\ .
$$
For the sake of simplicity, we shall focus our attention to the physically
most relevant case of central potentials, i. e., potentials which depend only
on the modulus $|{\bf x}|$ of the configuration-space relative coordinate:
\begin{equation}
V = V(|{\bf x}|)\ .
\label{eq:centralpot}
\end{equation}
In the above form, the spinless Salpeter equation appears to be a very
promising candidate for the (semi)relativistic description of hadrons as
bound states of (constituent) quarks within the framework of potential models
\cite{lucha91,lucha92,lucha95}, or, at least, the first step in the correct
direction \cite{gara89,lucha92sign}.

However, the presence of the relativistic kinetic-energy operator
(\ref{eq:kinenergy}) in (\ref{eq:ham-sseq}) or, to do justice to the spinless
Salpeter equation, the nonlocality of this operator $H$, that is, more
precisely, of either the kinetic-energy operator $T$ in configuration space
or the interaction-potential operator $V$ in momentum space, renders
difficult to arrive at rigorous analytical statements about the corresponding
energy spectrum. In view of this, numerous attempts to circumvent these
problems have been proposed. Some very brief account of the history of these
attempts may be found, for instance, in Ref.~\cite{lucha94}. These approaches
include, among others, the development of elaborate numerical approximation
methods \cite{nickisch84,lucha91num,fulcher93} as well as the construction of
effective Hamiltonians which, in spite of their apparently nonrelativistic
form, incorporate relativistic effects by sophisticated momentum dependence
of the involved parameters \cite{lucha93eff}. A lot of information on the
solutions of the spinless Salpeter equation may even be gained by application
of a relativistic virial theorem \cite{lucha89rvt}, most easily derived from
a rather general ``master virial theorem'' \cite{lucha90rvt}.

The (from the physical point of view perhaps most interesting) case of a
Coulomb-type static interaction potential, the so-called relativistic Coulomb
problem, has been investigated particularly carefully. For the corresponding
lowest-lying energy eigenvalues, both lower \cite{herbst77,martin89} and
upper \cite{martin89,lucha94varbound,lucha96rcprefer,lucha96rcplowly} bounds
have been derived and series expansions \cite{leyaouanc94} in powers of the
involved fine structure constant have been given.

Here, we intend to pave the way for the calculation of upper bounds on the
energy eigenvalues of the spinless Salpeter equation with rather arbitrary
interaction potentials. To this end, we apply the famous min--max
principle---which controls any such attempt---in a particular basis of our
trial space, characterized by generalized Laguerre polynomials.

\section{Minimum--Maximum Principle and Rayleigh--Ritz Variational
Technique}

The derivation of upper bounds on the eigenvalues of some operator $H$ makes,
of course, only sense for those operators $H$ which are bounded from below.
Accordingly, let us assume from now on that the arbitrary interaction
potential (\ref{eq:centralpot}) in our semirelativistic Hamiltonian
(\ref{eq:ham-sseq}) is such that this necessary prerequisite holds. For
example, for the crucial case of a Coulomb-type static interaction potential,
the so-called relativistic Coulomb problem, the demanded semi-boundedness of
the spectrum of the Hamiltonian $H$ has been (rigorously) demonstrated by
Herbst~\cite{herbst77}.

The theoretical basis as well as the primary tool for the derivation of
rigorous upper bounds on the eigenvalues of some self-adjoint operator is,
beyond doubt, the so-called min--max principle \cite{reed-simon}. An
immediate consequence of this min--max principle is the Rayleigh--Ritz
technique: Let $H$ be a semi-bounded self-adjoint operator. Let $E_k$, $k =
0,1,2,\dots$, denote the eigenvalues of $H$, ordered according to $E_0\le
E_1\le E_2\le\dots$. Let $D_d$ be some $d$-dimensional subspace of the domain
of $H$ and let $\widehat E_k$, $k = 0,1,\dots,d-1$, denote all $d$
eigenvalues of this operator $H$ restricted to the space $D_d$, ordered
according to $\widehat E_0\le\widehat E_1\le\dots\le\widehat E_{d-1}$. Then
the $k$th eigenvalue $E_k$ (counting multiplicity\footnote{\normalsize\ For
instance, for a Hamiltonian $H$ depending only on the moduli of momentum
${\bf p}$ and coordinate ${\bf x}$, respectively, states of given orbital
angular momentum but different projections of the latter will be
degenerate.}) of $H$ satisfies the inequality
$$
E_k \le \widehat E_k\ ,\quad k = 0,1,\dots,d-1\ .
$$
(For a discussion of the history of inequalities and variational methods for
eigenvalue problems, see, e.~g., Ref.~\cite{weinstein72}; for some
applications, see, e.~g., Ref.~\cite{flamm82}.)

Now, let us assume that this $d$-dimensional subspace $D_d$ is spanned by
some set of $d$ orthonormalized (and therefore beyond doubt linearly
independent) basis vectors $|\psi_k\rangle$, $k = 0,1,\dots,d-1$:
$$
\langle\psi_i|\psi_j\rangle = \delta_{ij}\ ,\quad i,j = 0,1,\dots,d-1\ .
$$
Then the set of eigenvalues $\widehat E$ may immediately be determined as the
$d$ roots of the characteristic equation
\begin{equation}
\det\left(\langle\psi_i|H|\psi_j\rangle - \widehat E\,\delta_{ij}\right) = 0\ ,
\quad i,j = 0,1,\dots,d-1\ ,
\label{eq:chareq}
\end{equation}
as becomes clear from an expansion of any eigenvector of the restricted
operator $H$ in terms of the set of basis vectors $|\psi_k\rangle$, $k =
0,1,\dots,d-1$, of the subspace $D_d$.

\section{Generalized Laguerre Basis}

The crucial step in any investigation of the present type is the suitable
choice of a basis in the subspace $D_d$. For the case of the semirelativistic
Hamiltonian (\ref{eq:ham-sseq}), we find it convenient to work in a basis
which involves the so-called generalized Laguerre polynomials. The latter are
specific orthogonal polynomials, defined by the power series \cite{abramow}
$$
L_k^{(\gamma)}(x) = \sum_{r=0}^k\,(-1)^r \left(\begin{array}{c}k+\gamma\\ k-r
\end{array}\right)\frac{x^r}{r!}
$$
and normalized according to \cite{abramow}
$$
\int\limits_0^\infty{\rm d}x\,x^\gamma\exp(-x)\,L_k^{(\gamma)}(x)\,
L_{k'}^{(\gamma)}(x) = \frac{\Gamma(\gamma+k+1)}{k!}\,\delta_{kk'}\ .
$$
Consequently, introducing two variational parameters, namely, one, $\mu$,
with the dimension of mass as well as a dimensionless one, $\beta$, a generic
trial vector $|\psi\rangle$ of the subspace $D_d$, with orbital angular
momentum $\ell$ and its projection $m$, will be characterized by the
following admittedly very suggestive ansatz for its coordinate-space
representation $\psi_{k,\ell m}({\bf x})$:
\begin{equation}
\psi_{k,\ell m}({\bf x})
= {\cal N}\,|{\bf x}|^{\ell+\beta-1}\exp(-\mu\,|{\bf x}|)\,
L_k^{(\gamma)}(2\,\mu\,|{\bf x}|)\,{\cal Y}_{\ell m}(\Omega_{\bf x})\ ,
\label{eq:ansatz}
\end{equation}
where normalizability restricts the variational parameter $\mu$ to positive
values,
$$
\mu>0\ .
$$
Here, ${\cal Y}_{\ell m}(\Omega)$ are the spherical harmonics for angular
momentum $\ell$ and projection $m$ depending on the solid angle $\Omega$;
they are orthonormalized according to
\begin{equation}
\int{\rm d}\Omega\,{\cal Y}^\ast_{\ell m}(\Omega)\,{\cal Y}_{\ell'm'}(\Omega)
= \delta_{\ell\ell'}\,\delta_{mm'}\ .
\label{eq:spharorth}
\end{equation}
The proper orthonormalization of the ansatz (\ref{eq:ansatz}) fixes the
parameter $\gamma$ necessarily to the value $\gamma = 2\,\ell+2\,\beta$ and
determines the normalization constant ${\cal N}$:
\begin{eqnarray*}
\psi_{k,\ell m}({\bf x}) &=& \sqrt{\frac{(2\,\mu)^{2\ell+2\beta+1}\,k!}
{\Gamma(2\,\ell+2\,\beta+k+1)}}\,|{\bf x}|^{\ell+\beta-1}
\exp(-\mu\,|{\bf x}|)\\[1ex]
&\times&L_k^{(2\ell+2\beta)}(2\,\mu\,|{\bf x}|)\,
{\cal Y}_{\ell m}(\Omega_{\bf x})
\end{eqnarray*}
satisfies the normalization condition
$$
\int{\rm d}^3x\,\psi_{k,\ell m}^\ast({\bf x)}\,\psi_{k',\ell'm'}({\bf x)}
= \delta_{kk'}\,\delta_{\ell\ell'}\,\delta_{mm'}\ .
$$
Rather obviously, normalizability constrains the variational parameter
$\beta$ too, namely, to a range characterized by $2\,\beta>-1$, i.~e., to the
range
$$
\beta > -\frac{1}{2}\ .
$$
The Fourier transform $\widetilde\psi_{k,\ell m}({\bf p})$ of the above trial
function involves the hypergeometric series $F$, defined with the help of the
gamma function $\Gamma$ by \cite{abramow}
$$
F(u,v;w;z) = \frac{\Gamma(w)}{\Gamma(u)\,\Gamma(v)}\,\sum_{n=0}^\infty\,
\frac{\Gamma(u+n)\,\Gamma(v+n)}{\Gamma(w+n)}\,\frac{z^n}{n!}\ ;
$$
it reads
\begin{eqnarray*}
\widetilde\psi_{k,\ell m}({\bf p})
&=& \sqrt{\frac{(2\,\mu)^{2\ell+2\beta+1}\,k!}
{\Gamma(2\,\ell+2\,\beta+k+1)}}\,\frac{(-i)^\ell\,|{\bf p}|^\ell}{2^{\ell+1/2}
\,\Gamma\left(\ell+\frac{3}{2}\right)}\\[1ex]
&\times& \sum_{r=0}^k\,\frac{(-1)^r}{r!}
\left(\begin{array}{c}k+2\,\ell+2\,\beta\\ k-r\end{array}\right)
\frac{\Gamma(2\,\ell+\beta+r+2)\,(2\,\mu)^r}
{({\bf p}^2+\mu^2)^{(2\ell+\beta+r+2)/2}}\\[1ex]
&\times& F\left(\frac{2\,\ell+\beta+r+2}{2},-\frac{\beta+r}{2};
\ell+\frac{3}{2};\frac{{\bf p}^2}{{\bf p}^2+\mu^2}\right)
{\cal Y}_{\ell m}(\Omega_{\bf p})
\end{eqnarray*}
and satisfies the normalization condition
$$
\int{\rm d}^3p\,\widetilde\psi_{k,\ell m}^\ast({\bf p)}\,
\widetilde\psi_{k',\ell'm'}({\bf p)}
= \delta_{kk'}\,\delta_{\ell\ell'}\,\delta_{mm'}\ .
$$

In principle, it is straightforward to calculate the expectation values
$$
H_{ij}\equiv\langle\psi_i|H|\psi_j\rangle
$$
of the Hamiltonian (\ref{eq:ham-sseq}), necessary for applying the min--max
principle. Due to the orthonormalization (\ref{eq:spharorth}) of the
spherical harmonics ${\cal Y}_{\ell m}(\Omega)$, however, only matrix
elements taken between states of identical orbital angular momentum $\ell$
and its projection $m$ will be nonvanishing.

\section{Power-Law Potentials}

When speculating about the possible shape of a physically meaningful (or
phenomenologically acceptable) interaction potential, the very first idea
which unavoidably comes to one's mind as a reasonable candidate is an
interaction potential of the power-law form, the power being only constrained
by requiring that the Hamiltonian is bounded from below:
\begin{equation}
V(|{\bf x}|) = \sum_n a_n\,|{\bf x}|^{b_n}\ ,
\label{eq:polapot}
\end{equation}
with sets of arbitrary real constants $a_n$ and $b_n$, the latter only
subject to the constraint
$$
b_n\ge -1\quad\mbox{if}\quad a_n<0\ .
$$
By close inspection of our ansatz (\ref{eq:ansatz}) it should become clear
that we are able to handle even potentials of the type
``power--times--exponential,'' that is, potentials of the form
$$
V(|{\bf x}|) = \sum_n a_n\,|{\bf x}|^{b_n}\exp(c_n\,|{\bf x}|)\ ,\quad
b_n\ge -1\quad\mbox{if}\quad a_n<0\ .
$$

It is a rather simple task to write down the matrix elements for the
power-law potential (\ref{eq:polapot}):
\begin{eqnarray*}
V_{ij} &\equiv& \langle\psi_i|V(|{\bf x}|)|\psi_j\rangle\\[1ex]
&=& \sum_n\,a_n\int{\rm d}^3x\,\psi_{i,\ell m}^\ast({\bf x})\,|{\bf x}|^{b_n}
\,\psi_{j,\ell m}({\bf x})\\[1ex]
&=& \sqrt{\frac{i!\,j!}{\Gamma(2\,\ell+2\,\beta+i+1)\,
\Gamma(2\,\ell+2\,\beta+j+1)}}\\[1ex]
&\times& \sum_n\,\frac{a_n}{(2\,\mu)^{b_n}}\,\sum_{r=0}^i\,\sum_{s=0}^j\,
\frac{(-1)^{r+s}}{r!\,s!}
\left(\begin{array}{c}i+2\,\ell+2\,\beta\\ i-r\end{array}\right)
\left(\begin{array}{c}j+2\,\ell+2\,\beta\\ j-s\end{array}\right)\\[1ex]
&\times& \Gamma(2\,\ell+2\,\beta+b_n+r+s+1)\ .
\end{eqnarray*}
For instance, considering merely radial excitations by letting $\ell=0$ and
choosing, just for the sake of definiteness, for the variational parameter
$\beta$ the value $\beta=1$, the explicit form of the potential matrix
$V\equiv (V_{ij})$ is
$$
V = \frac{1}{6}\,\sum_n\,\frac{a_n}{(2\,\mu)^{b_n}}\,\Gamma(3+b_n)
\left(\begin{array}{ccc}3&-\sqrt{3}\,b_n&\cdots\\
-\sqrt{3}\,b_n&3+b_n+b_n^2&\cdots\\ \vdots&\vdots&\ddots\end{array}\right)\ .
$$

\section{Analytically Evaluable Special Cases}

It should be really no great surprise that the evaluation of the matrix
elements of the kinetic-energy operator $T$,
\begin{eqnarray*}
T_{ij} &\equiv& \langle\psi_i|T({\bf p})|\psi_j\rangle\\[1ex]
&=& \int{\rm d}^3p\,\widetilde\psi_{i,\ell m}^\ast({\bf p})\,T({\bf p})\,
\widetilde\psi_{j,\ell m}({\bf p})\ ,
\end{eqnarray*}
is somewhat more delicate than the previous calculation of the matrix
elements of the power-law potentials $V$. Consequently, let us focus our
attention to those situations which allow for a fully analytic
evaluation of the above kinetic-energy matrix elements.

\subsection{Orbital Excitations}

On the one hand, we may restrict our formalism to the case $i=j=0$ but allow,
nevertheless, for still arbitrary values of the orbital angular momentum
$\ell$ (which means to consider arbitrary orbital excitations), and set
$\beta=1$. Then the matrix elements $V_{ij}$ of the power-law potential
(\ref{eq:polapot}) reduce to
$$
V_{00} = \frac{1}{\Gamma(2\,\ell+3)}\,\sum_n\,\frac{a_n}{(2\,\mu)^{b_n}}\,
\Gamma(2\,\ell+b_n+3)
$$
whereas for the matrix elements $T_{ij}$ of the kinetic energy
(\ref{eq:kinenergy}) we obtain
\begin{equation}
T_{00} = \frac{4^{\ell+2}\,[\Gamma(\ell+2)]^2}
{\sqrt{\pi}\,\Gamma\left(2\,\ell+\frac{7}{2}\right)}\,\mu\,
F\left(-\frac{1}{2},\ell+2;2\,\ell+\frac{7}{2};1-\frac{m^2}{\mu^2}\right)\ .
\label{eq:t00}
\end{equation}
At this point, our primary aim must be to get rid of the hypergeometric
series $F$ in the above intermediate result.
\begin{itemize}
\item In the ultrarelativistic limit, realized in the case of vanishing mass
$m$ of the involved particles, that is, for $m=0$, the hypergeometric series
$F$ in (\ref{eq:t00}) may be simplified with the help of the relation
\cite{abramow}
$$
F(u,v;w;1) = \frac{\Gamma(w)\,\Gamma(w-u-v)}{\Gamma(w-u)\,\Gamma(w-v)}
$$
for
$$
w\ne 0,-1,-2,\dots\ ,\quad \Re(w-u-v)>0\ ,
$$
in order to yield for the kinetic-energy matrix element $T_{00}$,
Eq.~(\ref{eq:t00}), the much more innocent expression
$$
T_{00} = \frac{2\,[\Gamma(\ell+2)]^2}
{\Gamma\left(\ell+\frac{3}{2}\right)\Gamma\left(\ell+\frac{5}{2}\right)}\,
\mu \ .
$$
The resulting upper bounds, $H_{00}$, can be optimized be minimizing $H_{00}$
with respect to the variational parameter $\mu$. For instance, for a linear
potential $V(|{\bf x}|) = a\,|{\bf x}|$, this minimization procedure thus
yields
$$
\min_{\mu>0}H_{00} = 2\,\Gamma(\ell+2)\,\sqrt{\frac{(2\,\ell+3)\,a}
{\Gamma\left(\ell+\frac{3}{2}\right)\Gamma\left(\ell+\frac{5}{2}\right)}}\ .
$$
In the limit of large orbital angular momenta $\ell$, that is, for
$\ell\to\infty$, this minimal upper bound turns out to be not in conflict
with the experimentally well-established linearity of ``Regge trajectories:''
$$
\lim_{\ell\to\infty}\left(\min_{\mu>0}H_{00}\right)^2 = 8\,a\,\ell\ ,
$$
which is in striking accordance with all previous findings
\cite{kang75,lucha91regge}.
\item Fixing the variational parameter $\mu$ to the particular value $\mu=m$
allows us to take advantage from the fact that
$$
F(u,v;w;0) =1\ ,
$$
whence the kinetic-energy matrix element $T_{00}$, Eq.~(\ref{eq:t00}),
reduces to
$$
T_{00} = \frac{4^{\ell+2}\,[\Gamma(\ell+2)]^2}
{\sqrt{\pi}\,\Gamma\left(2\,\ell+\frac{7}{2}\right)}\,m\ .
$$
\end{itemize}

\subsection{Radial Excitations}

On the other hand, considering only states of vanishing orbital angular
momentum $\ell$, i.~e., only states with $\ell=0$, confines our investigation
to the analysis of radial excitations. In this case, we may use the relation
\cite{abramow}
$$
F\left(u,1-u;\frac{3}{2};\sin^2z\right) =
\frac{\sin[(2\,u-1)\,z]}{(2\,u-1)\sin z}
$$
in order to recast the hypergeometric series $F$ in the momentum-space
representation $\widetilde\psi_{k,00}(|{\bf p}|)$ of our trial states into
the form
\begin{eqnarray*}
F\left(\frac{\beta+r+2}{2},-\frac{\beta+r}{2};\frac{3}{2};
\frac{{\bf p}^2}{{\bf p}^2+\mu^2}\right)
&=& \frac{\displaystyle\sqrt{{\bf p}^2+\mu^2}}{(\beta+r+1)\,|{\bf p}|}\\[1ex]
&\times& \sin\left[(\beta+r+1)\arctan\frac{|{\bf p}|}{\mu}\right]\ .
\end{eqnarray*}
Simplifying the momentum-space trial function $\widetilde\psi_{k,00}(|{\bf
p}|)$ in this way,
\begin{eqnarray*}
\widetilde\psi_{k,00}(|{\bf p}|) &=& \sqrt{\frac{k!}
{\mu\,\Gamma(2\,\beta+k+1)}}\,\frac{2^\beta}{\pi\,|{\bf p}|}\\[1ex]
&\times& \sum_{r=0}^k\,\frac{(-2)^r}{r!}\left(\begin{array}{c}k+2\,\beta\\
k-r\end{array}\right)\Gamma(\beta+r+1)\\[1ex]
&\times& \left(1+\frac{{\bf p}^2}{\mu^2}\right)^{-(\beta+r+1)/2}
\sin\left[(\beta+r+1)\arctan\frac{|{\bf p}|}{\mu}\right]\ ,
\end{eqnarray*}
the matrix elements $T_{ij}$ of the kinetic energy (\ref{eq:kinenergy})
immediately become
\begin{eqnarray*}
T_{ij} &=& \sqrt{\frac{i!\,j!}{\Gamma(2\,\beta+i+1)\,\Gamma(2\,\beta+j+1)}}\,
\frac{4^{\beta+1}}{\pi}\,\mu\\[1ex]
&\times& \sum_{r=0}^i\,\sum_{s=0}^j\,\frac{(-2)^{r+s}}{r!\,s!}
\left(\begin{array}{c}i+2\,\beta\\ i-r\end{array}\right)
\left(\begin{array}{c}j+2\,\beta\\ j-s\end{array}\right)\\[1ex]
&\times& \Gamma(\beta+r+1)\,\Gamma(\beta+s+1)\,I_{rs}\ ,
\end{eqnarray*}
where $I_{rs}$ denotes the only remaining integration,
\begin{eqnarray*}
I_{rs} &\equiv&
\int\limits_0^\infty{\rm d}y\,\sqrt{y^2+\frac{m^2}{\mu^2}}\\[1ex]
&\times& \frac{\cos[(r-s)\arctan y] - \cos[(2\,\beta+r+s+2)\arctan y]}
{(1+y^2)^{(2\beta+r+s+2)/2}}\ .
\end{eqnarray*}
This integration may, of course, always be performed by some standard
numerical integration procedure. However, for $\mu=m$, the integral $I_{rs}$
simplifies to
$$
I_{rs} = \int\limits_0^\infty{\rm d}y\,
\frac{\cos[(r-s)\arctan y] - \cos[(2\,\beta+r+s+2)\arctan y]}
{(1+y^2)^{(2\beta+r+s+1)/2}}\ ,
$$
which, for $2\,\beta$ integer and thus, because of the previous
normalizability constraint $2\,\beta>-1$, non-negative, i.~e., for the values
$2\,\beta=0,1,2,\dots$, may be evaluated with the help of the expansion
$$
\begin{array}{r}\cos(N\arctan y)
= \displaystyle\frac{1}{(1+y^2)^{N/2}}\,\displaystyle\sum_{n=0}^N
\left(\begin{array}{c}N\\n\end{array}\right)\cos\left(\frac{n\,\pi}{2}\right)
y^n\\[3ex]\mbox{for}\ N=0,1,2,\dots\ ,\end{array}
$$
with the result
\begin{eqnarray*}
I_{rs} &=& \frac{1}{2}
\left[\Gamma\left(\frac{2\,\beta+r+s+|r-s|+1}{2}\right)\right]^{-1}
\sum_{n=0}^{|r-s|}\left(\begin{array}{c}|r-s|\\ n\end{array}\right)\\[1ex]
&\times& \Gamma\left(\frac{n+1}{2}\right)
\Gamma\left(\frac{2\,\beta+r+s+|r-s|-n}{2}\right)
\cos\left(\frac{n\,\pi}{2}\right)\\[1ex]
&-& \frac{1}{2}\left[\Gamma\left(2\,\beta+r+s+\frac{3}{2}\right)\right]^{-1}
\sum_{n=0}^{2\beta+r+s+2}
\left(\begin{array}{c}2\,\beta+r+s+2\\ n\end{array}\right)\\[1ex]
&\times& \Gamma\left(\frac{n+1}{2}\right)
\Gamma\left(2\,\beta+r+s+1-\frac{n}{2}\right)
\cos\left(\frac{n\,\pi}{2}\right)\ .
\end{eqnarray*}
The case $\beta=0$, however, requires special care for the following reason.
For $\beta=0$, the integral $I_{00}$ and therefore also the kinetic-energy
matrix element $T_{00}$ become singular, as may be read off from the above
explicit expression for the integral $I_{rs}$. This singularity may be
cancelled by the contribution of a Coulomb-type term $\kappa\,|{\bf x}|^{-1}$
in the power-law potential (\ref{eq:polapot}) if the involved coupling
constant $\kappa$ takes some particular, ``critical'' value. This
cancellation can then be made manifest by observing that
\cite{lucha96rcplowly}
$$
\lim_{\beta\to 0}\,\int\limits_0^\infty{\rm d}y\,
\frac{1 - \cos[(2+2\,\beta)\arctan y]}{(1+y^2)^{1/2+\beta}}
= 2\lim_{\beta\to 0}\,\int\limits_0^\infty{\rm d}y\,
\frac{y^2}{(1+y^2)^{3/2+\beta}}\ .
$$
Explicitly, for $\beta=1$, the kinetic-energy matrix $T\equiv(T_{ij})$ is
given by
$$
T = \frac{128}{15\,\pi}\,m\left(\begin{array}{ccc}
1&\displaystyle\frac{\sqrt{3}}{7}&\cdots\\[2ex]
\displaystyle\frac{\sqrt{3}}{7}&\displaystyle\frac{11}{9}&\cdots\\[2ex]
\vdots&\vdots&\ddots\end{array}\right)\ .
$$

In any case, our approach yields analytic expressions for the matrix elements
$H_{ij}$ of our semirelativistic Hamiltonian $H$ with an interaction
potential out of the rather large class given by the power-law form
(\ref{eq:polapot}). In principle, the $d$ (real) roots of the characteristic
equation (\ref{eq:chareq}) may be determined algebraically up to and
including the case $d=4$, entailing, of course, analytic expressions of
rather rapidly increasing complexity. For larger values of the dimension $d$
of our trial space $D_d$, the resulting energy matrix, $(H_{ij})$, may be
easily diagonalized numerically, however, without the necessity to apply
time-consuming integration procedures.

In order to be able to estimate and appreciate the quality of all the upper
bounds obtained in this way, we apply the above results to four prototype
potentials, namely, to
\begin{itemize}
\item the harmonic-oscillator potential,
$$
V(|{\bf x}|) = \omega\,|{\bf x}|^2\ ,\quad\omega>0\ ,
$$
\item the Coulomb potential,
$$
V(|{\bf x}|) = -\frac{\kappa}{|{\bf x}|}\ ,\quad\kappa>0\ ,
$$
\item the linear potential,
$$
V(|{\bf x}|) = a\,|{\bf x}|\ ,\quad a>0\ ,
$$
and
\item the funnel potential,
$$
V(|{\bf x}|) = -\frac{\kappa}{|{\bf x}|} + a\,|{\bf x}|\ ,
\quad\kappa>0\ ,\quad a>0\ ,
$$
\end{itemize}
for typical values \cite{lucha92sign} of the involved coupling parameters
$\omega$, $\kappa$, and $a$. The upper bounds on the energy eigenvalues of
the lowest-lying radial excitations (1S, 2S, 3S, and 4S in usual
spectroscopic notation) for the harmonic-oscillator, Coulomb, linear, and
funnel potentials are shown in Tables~\ref{tab:osci} through
\ref{tab:funnel}, respectively; the upper bounds on the respective energy
eigenvalues of just the first orbital excitation (1P again in usual
spectroscopic notation) for the above potentials are listed in
Table~\ref{tab:1pstates}.
\normalsize
\begin{table}[htb]
\caption{Energy eigenvalues of the spinless Salpeter equation with
harmonic-oscillator potential $V(|{\bf x}|) = \omega\,|{\bf x}|^2$, for the
parameter values $\mu=m=1\;\mbox{GeV}$, $\omega=0.5\;\mbox{GeV}^3$,
$\beta=1$, and a size $d\times d$ of the energy matrix $(H_{ij})$. Numbers in
italics (for small matrix sizes) indicate analytically obtained results. All
eigenvalues are given in units of GeV.}
\label{tab:osci}
\begin{center}
\begin{tabular}{ccccc}
\hline\hline\\[-1.5ex]
State&$1\times 1$&$2\times 2$&$25\times 25$&Schr\"odinger\\[1ex]
\hline\\[-1.5ex]
1S&{\it 4.2162}&{\it 3.9276}&3.8249&3.8249\\
2S&---&{\it 8.1085}&5.7911&5.7911\\
3S&---&---&7.4829&7.4823\\
4S&---&---&9.0215&9.0075\\[1ex]
\hline\hline
\end{tabular}
\end{center}
\end{table}
\begin{table}[htb]
\caption[]{Energy eigenvalues of the spinless Salpeter equation with Coulomb
potential $V(|{\bf x}|) = -\kappa/|{\bf x}|$, for the parameter values
\cite{lucha92sign} $\mu=m=1\;\mbox{GeV}$, $\beta=1$, $\kappa = 0.456$, and
the size $d\times d$ of the energy matrix $(H_{ij})$. Numbers in italics (for
small matrix sizes) indicate analytically obtained results. All eigenvalues
are given in units of GeV.}
\label{tab:coulomb}
\begin{center}
\begin{tabular}{cccc}
\hline\hline\\[-1.5ex]
State&$1\times 1$&$2\times 2$&$25\times 25$\\[1ex]
\hline\\[-1.5ex]
1S&{\it 2.2602}&{\it 2.0539}&1.9450\\
2S&---&{\it 3.0702}&1.9868\\
3S&---&---&2.0015\\
4S&---&---&2.0238\\[1ex]
\hline\hline
\end{tabular}
\end{center}
\end{table}
\begin{table}[htb]
\caption[]{Energy eigenvalues of the spinless Salpeter equation with the
linear potential $V(|{\bf x}|) = a\,|{\bf x}|$, for the parameter values
\cite{lucha92sign} $\mu=m=1\;\mbox{GeV}$, $\beta=1$, $a =
0.211\;\mbox{GeV}^2$, and the size $d\times d$ of the energy matrix
$(H_{ij})$. Numbers in italics (for small matrix sizes) indicate
analytically obtained results. All eigenvalues are given in units of GeV.}
\label{tab:linear}
\begin{center}
\begin{tabular}{cccc}
\hline\hline\\[-1.5ex]
State&$1\times 1$&$2\times 2$&$20\times 20$\\[1ex]
\hline\\[-1.5ex]
1S&{\it 3.0327}&{\it 2.8034}&2.7992\\
2S&---&{\it 4.0767}&3.3629\\
3S&---&---&3.8079\\
4S&---&---&4.1905\\[1ex]
\hline\hline
\end{tabular}
\end{center}
\end{table}
\begin{table}[htb]
\caption[]{Energy eigenvalues of the spinless Salpeter equation with the
funnel potential $V(|{\bf x}|) = -\kappa/|{\bf x}| + a\,|{\bf x}|$, for the
parameter values \cite{lucha92sign} $\mu=m=1\;\mbox{GeV}$, $\beta=1$, $\kappa
= 0.456$, $a = 0.211\mbox{ GeV}^2$, and the size $d\times d$ of the energy
matrix $(H_{ij})$. Numbers in italics (for small matrix sizes) indicate
analytically obtained results. All eigenvalues are given in units of GeV.}
\label{tab:funnel}
\begin{center}
\begin{tabular}{cccc}
\hline\hline\\[-1.5ex]
State&$1\times 1$&$2\times 2$&$20\times 20$\\[1ex]
\hline\\[-1.5ex]
1S&{\it 2.5767}&{\it 2.5182}&2.5162\\
2S&---&{\it 3.4499}&3.1570\\
3S&---&---&3.6337\\
4S&---&---&4.0348\\[1ex]
\hline\hline
\end{tabular}
\end{center}
\end{table}
\begin{table}[htb]
\caption[]{Energy eigenvalues for the 1P states of the spinless Salpeter
equation with the harmonic-oscillator potential $V(|{\bf x}|) = \omega\,|{\bf
x}|^2$, the Coulomb potential $V(|{\bf x}|) = -\kappa/|{\bf x}|$, the linear
potential $V(|{\bf x}|) = a\,|{\bf x}|$, and the funnel potential $V(|{\bf
x}|) = -\kappa/|{\bf x}| + a\,|{\bf x}|$, respectively, for the parameter
values \cite{lucha92sign} $\mu=m=1\;\mbox{GeV}$, $\beta=1$,
$\omega=0.5\;\mbox{GeV}^3$, $\kappa = 0.456$, $a = 0.211\;\mbox{GeV}^2$, and
the size $d\times d$ of the energy matrix $(H_{ij})$. Numbers in italics (for
small matrix sizes) indicate analytically obtained results. All eigenvalues
are given in units of GeV.}
\label{tab:1pstates}
\begin{center}
\begin{tabular}{ccccc}
\hline\hline\\[-1.5ex]
Potential&$1\times 1$&$20\times 20$&Schr\"odinger\\[1ex]
\hline\\[-1.5ex]
Harmonic oscillator&{\it 6.5094}&4.9015&4.9015\\
Coulomb&{\it 2.5314}&1.9875&---\\
Linear&{\it 3.2869}&3.1414&---\\
Funnel&{\it 3.0589}&2.9816&---\\[1ex]
\hline\hline
\end{tabular}
\end{center}
\end{table}
\large

For the case of the harmonic-oscillator potential, the corresponding
Hamiltonian $H$ in its momentum-space representation is equivalent to a
nonrelativistic Hamiltonian with some effective interaction potential, which
clearly is reminiscent of that troublesome square-root operator. In this
form, it is then rather easily accessible to numerical procedures for solving
a nonrelativistic Schr\"odinger equation \cite{falk85}. For comparison, we
quote, in Tables~\ref{tab:osci} and \ref{tab:1pstates}, the eigenvalues
obtained along these lines.

We find a very encouraging, rapid convergence of the upper bounds.

\section{Summary}

By application of the well-known min--max principle, which represents the
theoretical foundation of any computation of upper bounds on the eigenvalues
of self-adjoint operators, to trial spaces spanned by sets of basis states
which enable us to handle the square-root operator of the relativistic
kinetic energy $T$ in a satisfactory manner, we demonstrated how to derive
(for lowest-lying states even analytically!) upper bounds on the energy
levels of the spinless Salpeter equation with some (linear combination of)
power-law potentials. Interestingly, in the case of the funnel potential,
which is the prototype of almost all of the ``realistic,'' that is,
phenomenologically acceptable, inter-quark potentials used for the
description of hadrons as bound states of (constituent) quarks, the obtained
lowest-order approximation to the upper bound on, e.~g., the ground-state
energy is merely some 2 \% above the corresponding value. Of course, all the
bounds derived here may be improved numerically by a minimization with
respect to the variational parameters introduced.

\end{document}